\documentclass[12pt]{amsart}
\usepackage[utf8]{inputenc}
\usepackage[margin=1in, includehead, includefoot]{geometry}
\usepackage{amsthm,amssymb,amsmath}
\usepackage{enumitem}
\usepackage{mathrsfs}
\usepackage{mathtools}
\usepackage{fancyhdr}
\usepackage{indentfirst}
\usepackage{graphicx}
\usepackage{placeins}
\usepackage{wrapfig}

\newtheorem{lemma}{Lemma}[section]
\newtheorem{theorem}{Theorem}[section]
\newtheorem{corollary}{Corollary}[section]
\newtheorem{observation}{Observation}[section]
\newtheorem{prop}{Proposition}[section]
\newtheorem{conjecture}{Conjecture}

\DeclarePairedDelimiter\ceil{\lceil}{\rceil}

\newcommand{\toroman}[1]{\textit{\expandafter{\romannumeral #1\relax}}}

\newcommand{\cbeginproof}[0]{\par\noindent\textit{Proof.} }
\newcommand{\cendproof}[0]{ \qed\par\vspace{1em}}

\newcommand{\npcompleteproblem}[3]{ \par\vspace{0.5em}\noindent{\textbf{#1}}\newline \textbf{INSTANCE: } #2 \newline \textbf{QUESTION: } #3 \par\vspace{0.5em} }

\newcommand{\unlabeledsection}[1]{\par\vspace{1 em}\noindent\textbf{#1}\par\vspace{0.2em}}

\title{Fault-Tolerant Locating-Dominating sets with Error-correction}
\author{
    \small Devin C. Jean\\
    \small Computer Science Department \\
    \small Vanderbilt University\\
    \small \texttt{devin.c.jean@vanderbilt.edu}
    \\\\\and\\\\
    \small Suk J. Seo\\
    \small Computer Science Department\\
    \small Middle Tennessee State University\\
    \small \texttt{Suk.Seo@mtsu.edu}
}
\date{}

\begin{document}
\maketitle
\thispagestyle{empty}
\begin{abstract}
A \emph{locating-dominating} set is a subset of vertices representing ``detectors" in a graph $G$; each detector monitors its closed neighborhood and can distinguish its own location from its neighbors, and given all sensor input, the system can locate an ``intruder" anywhere in the graph.
We explore a fault-tolerant variant of locating-dominating sets, \emph{error-correcting locating-dominating} (ERR:LD) sets, which can tolerate an incorrect signal from a single detector.
In particular, we characterize error-correcting locating-dominating sets, and derive its existence criteria.
We also prove that the problem of determining the minimum cardinality of ERR:LD set in arbitrary graphs is NP-complete.
Additionally, we establish lower and upper bounds for the minimum density of ERR:LD sets in infinite grids and cubic graphs, and prove the lower bound for cubic graphs is sharp.
\end{abstract}

\vspace{1em}


\section{Introduction}\label{sec:intro}

Consider a system which uses a set of sensors or detectors placed at some subset of locations to automatically detect any ``intruder" in a facility or network.
Conceptually, an ``intruder" can be a real, physical intruder or some undesirable event, such as an error in a networked system.
These detection systems can be modeled as a graph using a subset of vertices as detectors.
Let $G$ be a graph with vertices $V(G)$ and edges $E(G)$.
The \emph{open-neighborhood} of a vertex $v \in V(G)$, denoted $N(v)$, is the set of all vertices adjacent to $v$: $N(v) = \{w \in V(G) : vw \in E(G)\}$.
The \emph{closed-neighborhood} of a vertex $v \in V(G)$, denoted $N[v]$, is the set of all vertices adjacent to $v$, as well as $v$ itself: $N[v] = N(v) \cup \{v\}$.

Of interest are detection systems capable of pinpointing an intruder's exact location by placing detectors at some subset of positions in the network or facility. 
The number and placement of detectors required to construct a detection system vary depending on the sensors' capabilities. 
Many detection systems modeled as a graph consider a detector that can identify an intruder only in its open or closed neighborhood.

A vertex set $S \subseteq V(G)$ is an \textit{identifying code} (IC) \cite{watching-sys, karpovsky} if for each vertex $v \in V(G)$ we have  $S \cap N[v] \neq \varnothing$ (that is, S is a \textit{dominating set}) and for any pair of distinct vertices $v$ and $u$ in $V(G)$ we have $N[v] \cap S \neq N[u] \cap S$; 
a vertex set $S \subseteq V(G)$ is an \textit{open-locating-dominating set} (OLD set) \cite{ourtri, old, oldtree} if for each vertex $v \in V(G)$ we have  $S \cap N(v) \neq \varnothing$ (that is, S is an\textit{ open-dominating se}t) and for any pair of distinct vertices $v$ and $u$ in $V(G)$ we have $N(v) \cap S \neq N(u) \cap S$.
In other words, an \emph{identifying code detector} at vertex $v$ can detect an intruder  in $N[v]$  and an \emph{open-locating-dominating detector} can detect an intruder at $N(v)$; thus, IC and OLD detectors can detect an intruder in the closed or open neighborhoods, respectively.

In this paper, we consider \emph{locating-dominating} (LD) detectors, which detect an intruder in $N[v]$, but can also tell if the intruder is at $v$ itself or somewhere in $N(v)$.
As introduced by Slater \cite{dom-ref-sets, dom-loc-acyclic}, a vertex set $S \subseteq V(G)$ is a \textit{locating-dominating set} if for each vertex $v \in V(G)$ we have  $S \cap N[v] \neq \varnothing$ and for any pair of distinct vertices $v$ and $u$ in $V(G)-S$ we have $N[v] \cap S \neq N[u] \cap S$. 
Lobstein \cite{dombib} maintains a bibliography of related parameters, which currently has over 440 papers.

Because LD detectors can distinguish their own location, LD sets are, in general, smaller than IC and OLD sets.
However, in real-world applications we often want some level of fault-tolerance to be built into the system; thus, several fault-tolerant variants of LD sets have been created.
One such variant is the \emph{redundant locating-dominating} (RED:LD) set, which is an LD set that can tolerate at most one detector being removed or going offline \cite{redld}.
Another variant is an \emph{error-detecting locating-dominating} (DET:LD) sets \cite{detld,ftld}, which can tolerate at most one false negative.
In this paper, we will explore a third, strongest variant called an \emph{error-correcting locating-dominating} (ERR:LD) set, which can tolerate one false negative or false positive; thus, it can correct any one error.
More general types of detector-based fault-tolerance have also been studied by Seo and Slater \cite{ftsets, gen}.

When constructing these detection systems, obviously we are interested in minimizing the size of subsets of vertices or the number of detectors. 
The \textit{locating-dominating number} LD(G) is the
minimum cardinality of an LD set for a graph $G$.
Similarly, the \textit{redundant-locating-dominating number} RED:LD(G), the \textit{error-detecting-locating-dominating number} DET:LD(G), and the \textit{error-correcting-locating-dominating number} ERR:LD(G) are the
minimum cardinalities of redundant-locating-dominating, error-detecting-locating-dominating, and error-correcting-locating-dominating sets, respectively, for $G$.

\begin{figure}[ht]
    \centering
    \begin{tabular}{c@{\hskip 3em}c@{\hskip 3em}c@{\hskip 3em}c}
        \includegraphics[width=0.2\textwidth]{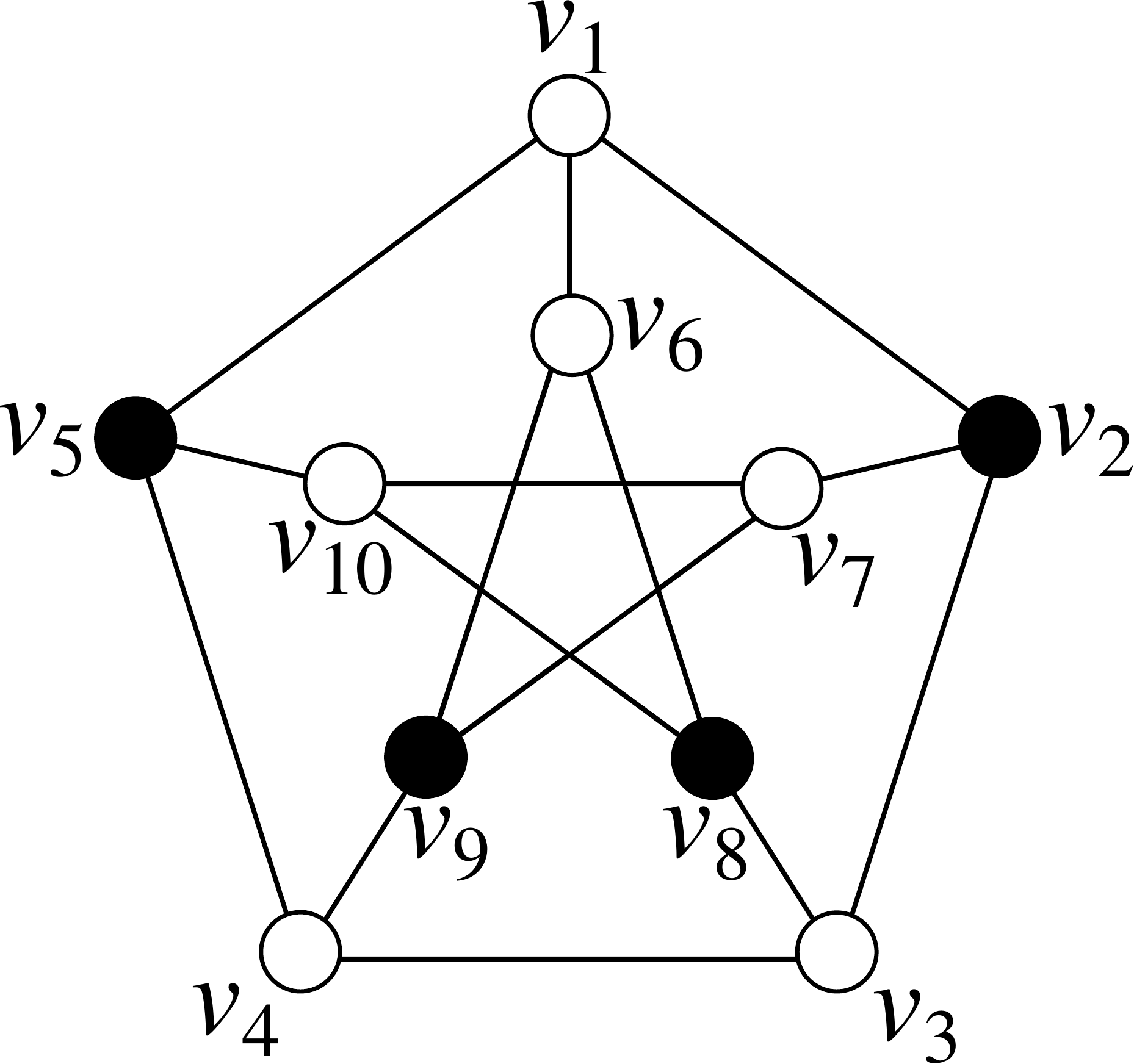} & \includegraphics[width=0.2\textwidth]{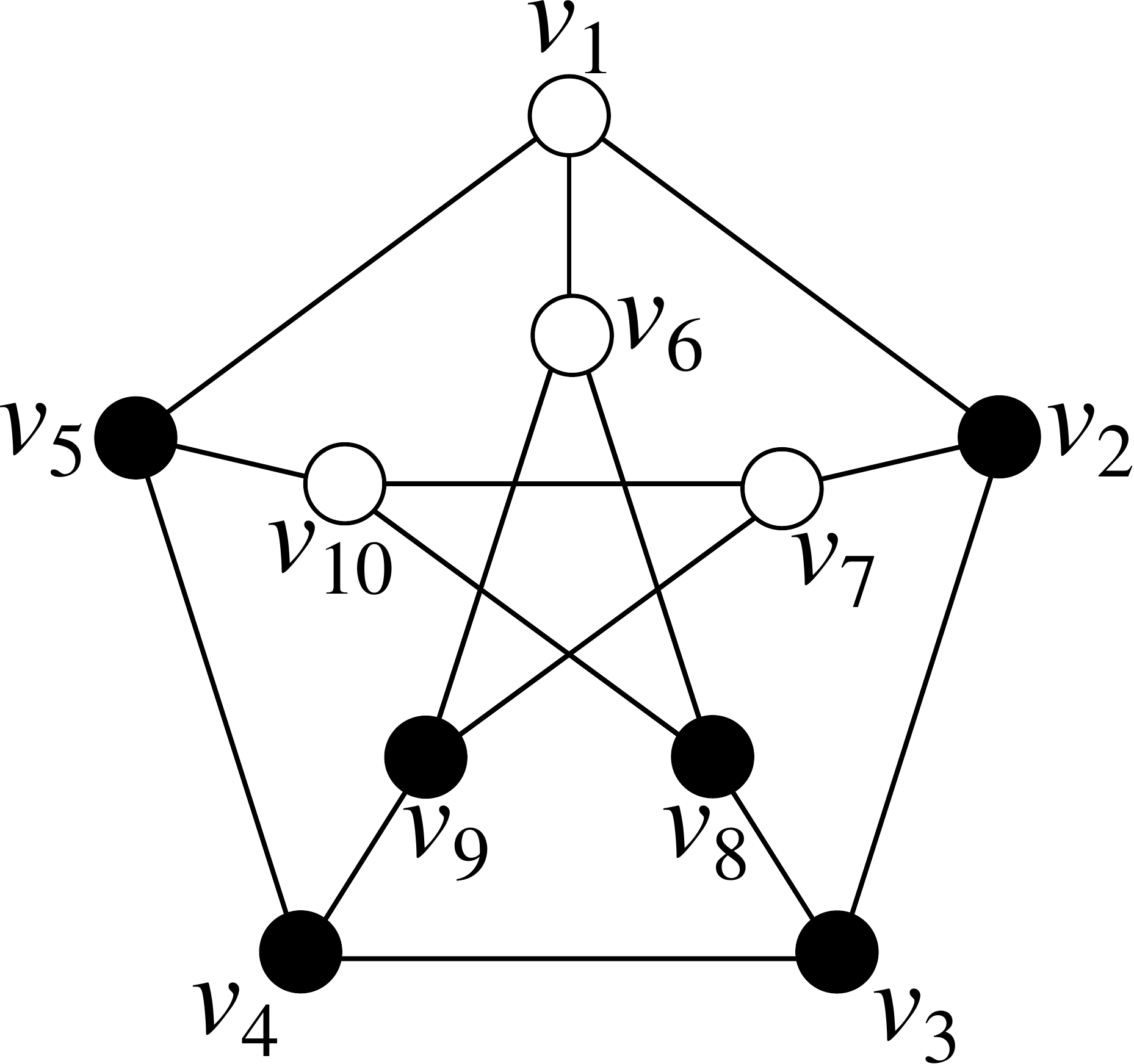} & \includegraphics[width=0.165\textwidth]{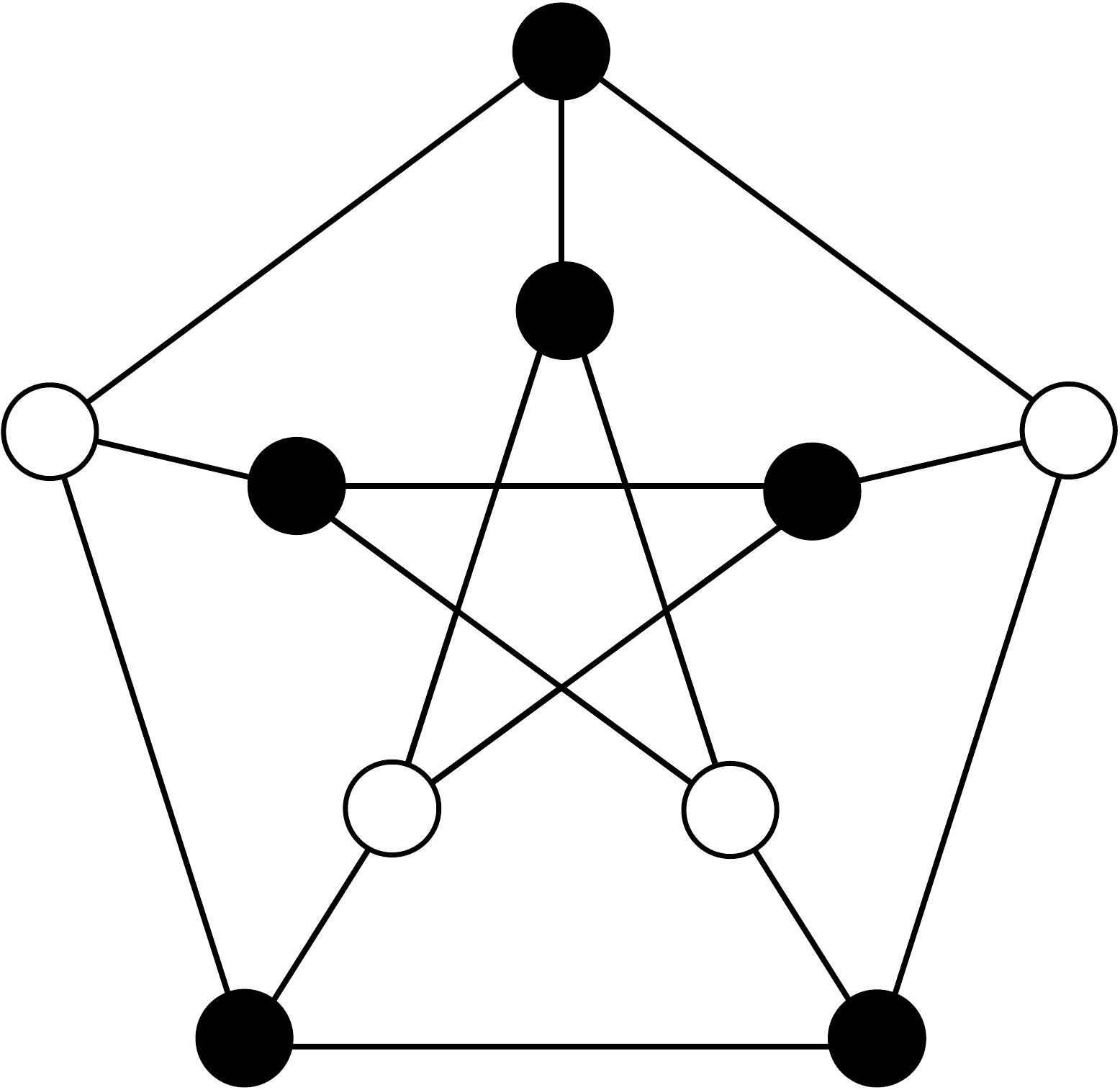} & \includegraphics[width=0.165\textwidth]{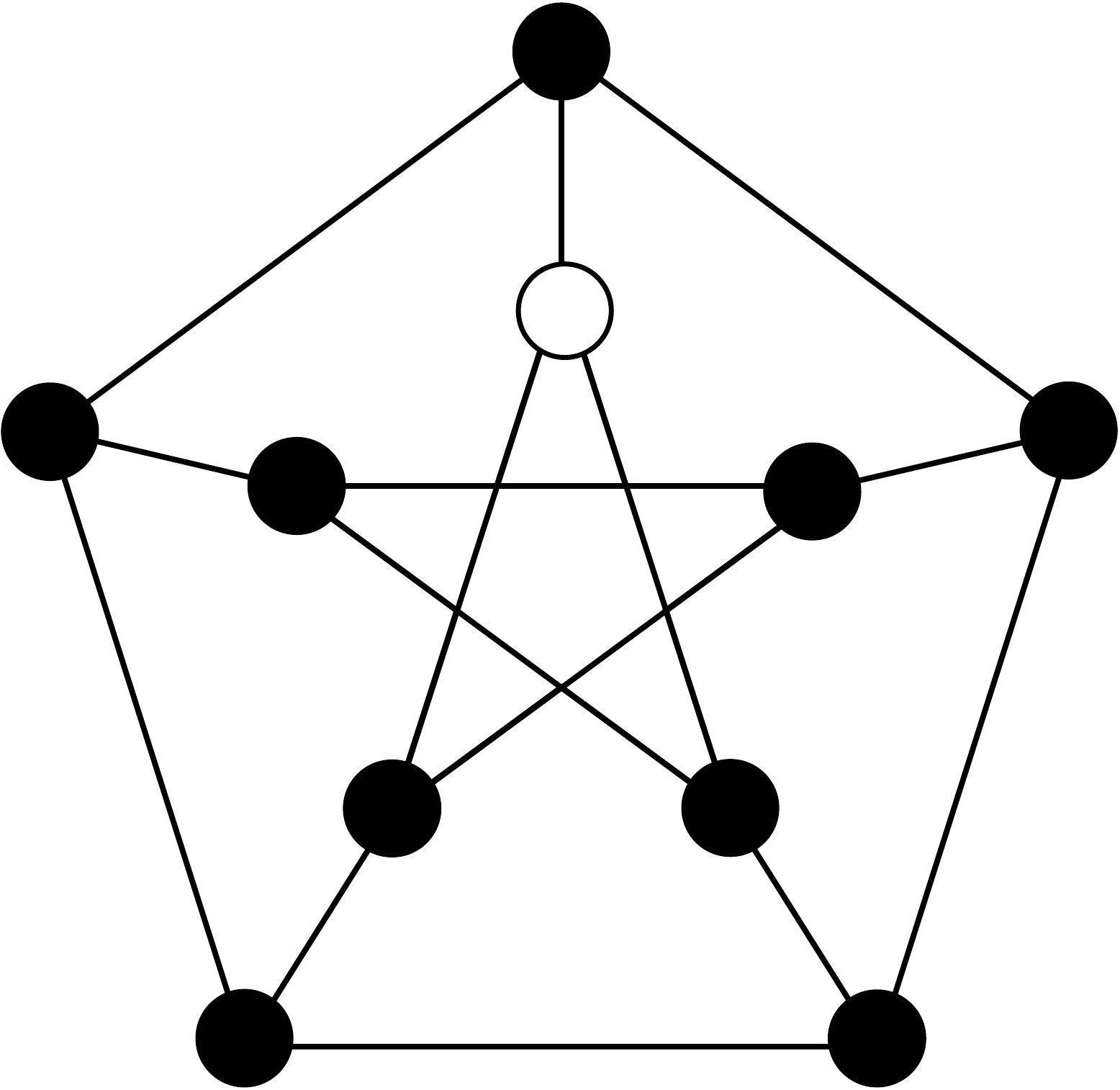} \\
        (a) & (b) & (c) & (d)
    \end{tabular}
    \caption{Optimal LD (a), RED:LD (b), DET:LD (c), and ERR:LD (d) sets on the Petersen graph.}
    \label{fig:petersen-examples}
\end{figure}

For example, Figure~\ref{fig:petersen-examples} shows LD, RED:LD, DET:LD, and ERR:LD sets of the minimum cardinality on the Petersen graph, $G$, hence we have $\textrm{LD}(G) = 4$, $\textrm{RED:LD}(G) = 6$, $\textrm{DET:LD}(G) = 6$, and $\textrm{ERR:LD}(G) = 9$.
We can verify that in Figure~\ref{fig:petersen-examples}~(a), the shaded vertices make up an LD set by checking each of the six non-detectors is dominated by a distinct subset of detectors. 
For example, we see a non-detector vertex $v_{1}$ is dominated by detectors at $v_{2}$ and $v_5$ and another non-detector vertex $v_6$ is dominated by detectors at $v_8$ and $v_9$.  
In Figure~\ref{fig:petersen-examples}~(b), the shaded vertex set $S = \{v_2, v_3, v_4, v_5, v_8, v_9\}$  makes up a RED:LD set because for any detector vertex $w \in S$, $S- \{w\}$ remains an LD set.
For the verification of the DET:LD set in Figure~\ref{fig:petersen-examples}~(c), refer to \cite{detld}.
The verification details for an ERR:LD set will be established in Section~\ref{sec:char}; for the verification of Figure~\ref{fig:petersen-examples}~(d), see the various examples throughout this paper.

In Section~\ref{sec:char} we characterize ERR:LD sets in general graphs and establish existence criteria.
In Section~\ref{sec:np-complete} we prove that the problem of determining $\textrm{ERR:LD}(G)$ for arbitrary graphs is NP-complete.
We then determine ERR:LD values for special classes of graphs, including cubic graphs and infinite grids, and conclude with future directions.

\section{Characterization}\label{sec:char}
In this section, we present the necessary and sufficient properties for an ERR:LD set in an arbitrary graph. 
Unlike an LD set, not all graphs have an ERR:LD set, so we also establish some existence criteria for ERR:LD sets.

For reference, we first include characterizations of LD sets, RED:LD sets, and DET:LD sets from previous research. 
Slater \cite{dom-loc-acyclic} introduced and presented properties of an LD set, and Jean and Seo \cite{redld, detld} established the following two theorems that characterize RED:LD sets and DET:LD sets.


\begin{theorem}[\cite{redld}]\label{theo:ld-char}
A set $S \subseteq V(G)$ is an LD set if and only if the following are true:
\begin{enumerate}[noitemsep, label=\roman*.]
    \item $\forall v \in V(G)-S$, $|N(v) \cap S| \ge 1$
    \item $\forall v,u \in V(G)-S$ with $v \neq u$, $|(N(v) \cap S) \triangle (N(u) \cap S)| \ge 1$
\end{enumerate}
\end{theorem}

\begin{theorem}[\cite{redld}]\label{theo:red-ld-char}
A set $S \subseteq V(G)$ is a RED:LD set if and only if the following are true:
\begin{enumerate}[noitemsep, label=\roman*.]
    \item $\forall v \in V(G)$, $|N[v] \cap S| \ge 2$
    \item $\forall v \in S$ and $\forall u \in V(G)-S$, $|((N(v) \cap S) \triangle (N(u) \cap S)) - \{v\}| \ge 1$
    \item $\forall v,u \in V(G)-S$ with $u \neq v$, $|(N(v) \cap S) \triangle (N(u) \cap S)| \ge 2$
\end{enumerate}
\end{theorem}

Slater \cite{ftld} proved several properties of DET:LD sets, and Jean and Seo \cite{detld} fully characterized DET:LD sets as follows.

\begin{theorem}[\cite{detld}]\label{theo:det-ld-char}
A set $S \subseteq V(G)$ is a DET:LD set if and only if the following are true:
\small{\begin{enumerate}[noitemsep, label=\roman*.]
    \item $\forall v \in V(G)$, $|N[v] \cap S| \ge 2$
    \item $\forall v,u \in S$ with $u \neq v$, $|(N(v) \cap S) \triangle (N(u) \cap S)| \ge 1$.
    \item $\forall v \in V(G)-S$ and $\forall u \in S$, $|(N(v) \cap S) - (N(u) \cap S)| \ge 2$ or $|(N(u) \cap S) - (N(v) \cap S)| \ge 1$
    \item $\forall v,u \in V(G)-S$ with $u \neq v$, $|(N(v) \cap S) - (N(u) \cap S)| \ge 2$ or $|(N(u) \cap S) - (N(v) \cap S)| \ge 2$
\end{enumerate}}
\end{theorem}

From Theorems \ref{theo:red-ld-char} and \ref{theo:det-ld-char}, we see that every DET:LD set is a RED:LD set.

\begin{corollary}\label{cor:det-is-red}
Every DET:LD set is also a RED:LD set.
\end{corollary}

\unlabeledsection{Characterization of ERR:LD sets}

The characterizations of various fault-tolerant LD sets require different levels of domination; for instance, LD sets require all vertices be dominated at least once, and RED:LD and DET:LD sets, at least twice.
For any fault-tolerant LD set variant $S \subseteq V(G)$, a vertex $v \in V(G)$ is \emph{$k$-dominated} if $|N[v] \cap S| = k$.
For example, if $S$ is a RED:LD set, then $\forall v \in V(G)$, $v$ is at least 2-dominated.
We say that two vertices are \emph{distinguished} if the system can eliminate one of them as the intruder location.
An ERR:LD set must have all vertex pairs be distinguished in order to find the intruder.

\begin{theorem}\label{theo:err-ld-char}
A detector set $S \subseteq V(G)$ is an ERR:LD set if and only if the following are true:
\begin{enumerate}[noitemsep, label=\roman*.]
    \item $\forall v \in V(G)$, $|N[v] \cap S| \ge 3$
    \item $\forall v,u \in S$ with $u \neq v$, $|((N(v) \cap S) \triangle (N(u) \cap S)) - \{v, u\}| \ge 1$
    \item $\forall v \in V(G)-S$ and $\forall u \in S$, $|((N(v) \cap S) \triangle (N(u) \cap S)) - \{u\}| \ge 2$
    \item $\forall v,u \in V(G)-S$ with $u \neq v$, $|(N(v) \cap S) \triangle (N(u) \cap S)| \ge 3$
\end{enumerate}
\end{theorem}
\begin{proof}
See Section~\ref{sec:proofs}.
\end{proof}

Note that the casing used in the converse of the proof for Theorem~\ref{theo:err-ld-char} can be used as a simple elimination algorithm for locating an intruder in the graph given all of the sensor values.

\vspace{0.5em}
The following corollary is provided as a mnemonic for the distinguishing requirements of ERR:LD sets.

\begin{corollary}\label{cor:err-ld-char}
The properties given in Theorem~\ref{theo:err-ld-char} can be rewritten as the following:
\begin{enumerate}[noitemsep, label=\roman*.]
    \item $\forall v \in V(G)$, $|N[v] \cap S| \ge 3$
    \item $\forall v,u \in V(G)$ with $u \neq v$, $|((N(v) \cap S) \triangle (N(u) \cap S)) - \{v, u\}| \ge 3 - |\{v,u\} \cap S|$
\end{enumerate}
\end{corollary}

Based on the properties of Theorem~\ref{theo:err-ld-char} (but more clearly seen in Corollary~\ref{cor:err-ld-char}), for any ERR:LD set $S \subseteq V(G)$, we say that two vertices $u,v \in V(G)$ are \emph{$k$-distinguished} if $|((N(u) \cap S) \triangle (N(v) \cap S)) - \{u,v\}| \ge k$.

\unlabeledsection{Example of using Theorem~\ref{theo:err-ld-char}}

\begin{figure}[ht]
    \centering
    \includegraphics[width=0.35\textwidth]{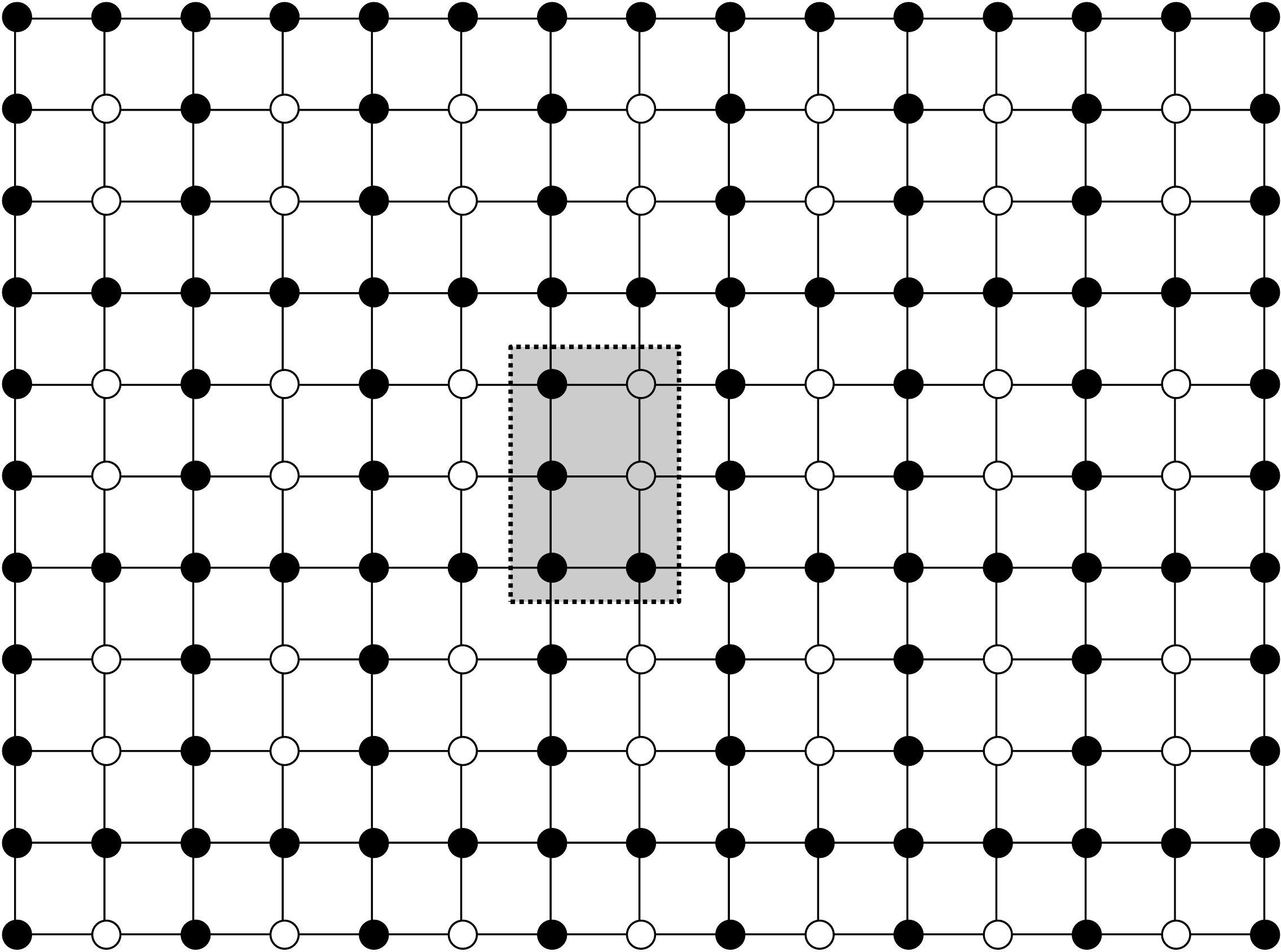}
    \caption{An ERR:LD set on SQ. Shaded vertices denote detectors.}
    \label{fig:err-ld-sq-soln}
\end{figure}

\begin{wrapfigure}[11]{r}{0.3\textwidth}
    \centering
    \includegraphics[width=0.25\textwidth]{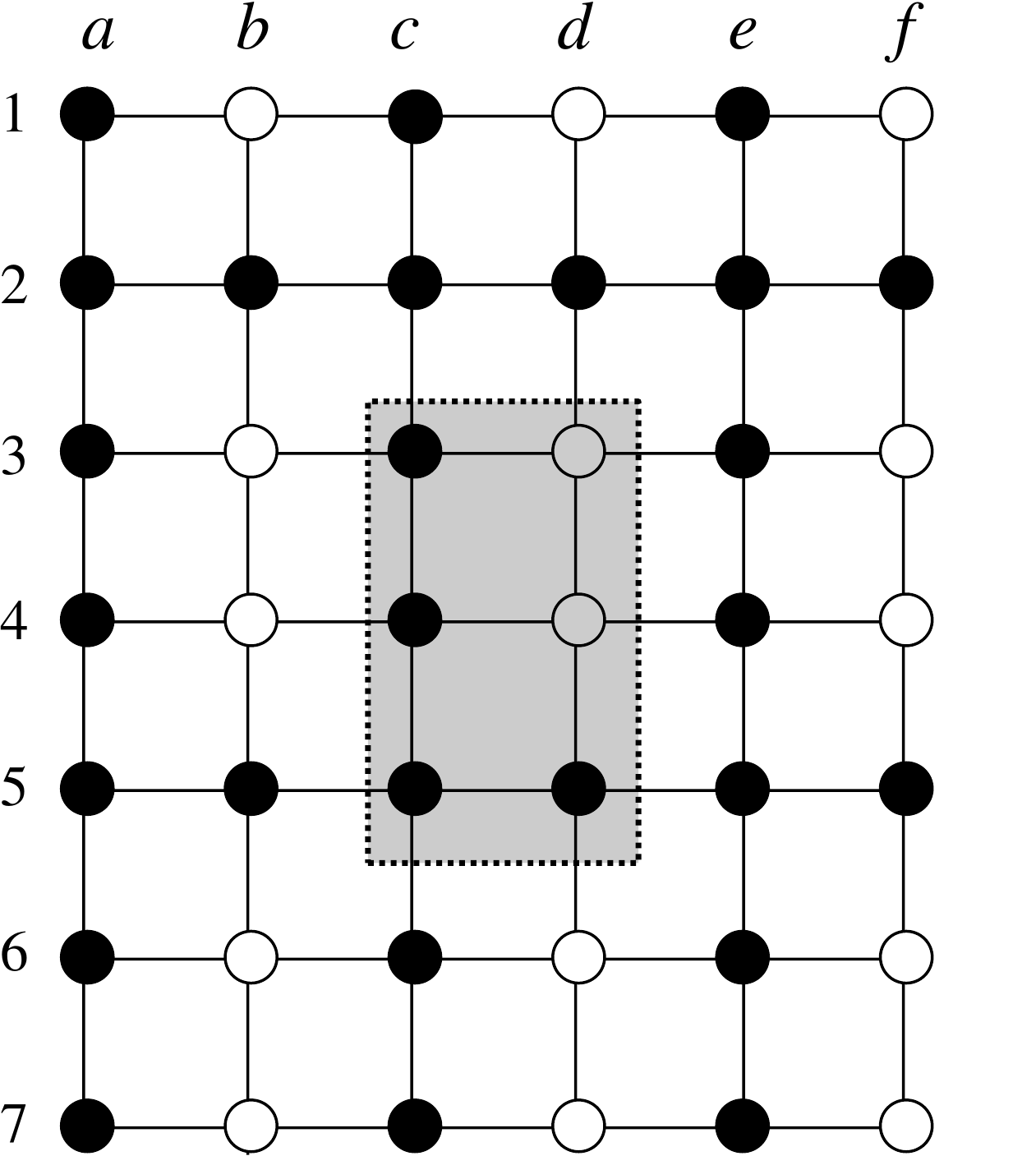}
    \caption{\\ERR:LD set}
    \label{fig:err-ld-one-tile}
\end{wrapfigure}

Let $S$ be the set of shaded vertices in Figure~\ref{fig:err-ld-sq-soln} on the \emph{infinite square grid}, SQ.
We will use the characterization given in Theorem~\ref{theo:err-ld-char} to verify that $S$ is an ERR:LD set for SQ.
Firstly, every vertex is at least 3-dominated, meeting the requirement in Theorem~\ref{theo:err-ld-char} \toroman{1}.
Next, we will show that all detector pairs, non-detector pairs, and mixed pairs (detector and non-detector) meet the requirements of Theorem~\ref{theo:err-ld-char} \toroman{2}-\toroman{4}.

Take the detector pair $(v_{c3}, v_{c4})$: aside from themselves, they are dominated by $v_{c2}$ and $v_{c5}$, and neither is common to both.
Thus, $v_{c3}$ and $v_{c4}$ are 2-distinguished; they need only be 1-distinguished, so they are distinguished.
As another example, take the detector pair $(v_{d5},v_{e4})$: aside from themselves, they are dominated by $v_{c5}$, $v_{e3}$,  and $v_{e5}$, and $v_{e5}$ is common to both.
Thus, $v_{d5}$ and $v_{e4}$ are 2-distinguished.
We observe that every pair of detectors is dominated by at most one common detector excluding themselves, therefore they are at least 2-distinguished, meeting the requirement in Theorem~\ref{theo:err-ld-char}~\toroman{2}.

Next, consider the non-detector pair $(v_{d3},v_{d4})$; we see they are 6-distinguished by $v_{b3}$, $v_{d2}$, $v_{e3}$, $v_{c4}$, $v_{d5}$, and $v_{e4}$; they need only be 3-distinguished, so they are distinguished.
We observe that every pair of non-detectors is dominated by at most one common detector, therefore they are at least 4-distinguished, meeting the requirement in Theorem~\ref{theo:err-ld-char}~\toroman{3}.

Lastly, take the mixed pair $(v_{c2},v_{d3})$; we find they are 3-distinguished by $v_{b2}$, $v_{c1}$, and $v_{e3}$, they need only be 2-distinguished, so they are distinguished.
As another example, we see the mixed pair $(v_{b4},v_{c3})$ is commonly dominated by $v_{c4}$, but it is distinguished by $v_{a4}$, $v_{b5}$, and $v_{c2}$.
We observe that every mixed pair is dominated by at most two common detectors and when they do have exactly two common dominators, the detector is 5-dominated; therefore they are at least 3-distinguished, meeting the requirement in Theorem~\ref{theo:err-ld-char}~\toroman{4}.
It can be shown that all pairs are distinguished, so $S$ is an ERR:LD set.

\unlabeledsection{Existence criteria for ERR:LD sets}

From Theorem~\ref{theo:det-ld-char}, we see that the weakest distinguishing requirement is between two detector vertices, which we can use to form existence criteria for ERR:LD sets.
As Foucaud et al. \cite{twin-free} described, two distinct vertices $u$ and $v$ of a graph are twins if $N[u] = N[v]$ or $N(u) = N(v)$.
A graph is \emph{twin-free} if it has no twins.

\begin{theorem}\label{theo:err-ld-exist}
An ERR:LD set exists if and only if $\delta(G) \ge 2$ and $G$ is twin-free.
\end{theorem}
\begin{proof}
Consider the detector set $S = V(G)$; then there are no non-detector vertices, meaning $S$ satisfies Theorem~\ref{theo:err-ld-char} properties \toroman{3} and \toroman{4}.
$S$ also satisfies property~\toroman{1} because $\forall v \in V(G)$, $|N[v] \cap S| = |N[v]| = 1 + |N(v)| \ge 3$.
Let $u,v \in V(G)$.
Suppose $uv \in E(G)$; we know $N[u] \neq N[v]$ because $G$ is twin-free, so $\exists w \in N[u] \triangle N[v] = (N(u) \triangle N(v)) - \{u,v\}$.
Otherwise, $uv \notin E(G)$; we know $N(u) \neq N(v)$ because $G$ is twin-free, so $\exists w \in N(u) \triangle N(v) = (N(u) \triangle N(v)) - \{u,v\}$.
In any case, we see that $w$ distinguishes $u$ and $v$, so $S$ also satisfies property~\toroman{2}.
Therefore, $S$ is an ERR:LD set on $G$.

For the converse, suppose $\delta(G) \le 1$ or $\exists v,u \in V(G)$ where $u$ and $v$ are twins.
If $\delta(G) \le 1$ then $\exists w \in V(G)$ such that $|N[w]| \le 2$, implying $w$ cannot be 3-dominated; this violates property~\toroman{1}, so there exists no ERR:LD set on $G$.
Otherwise, we assume $\exists v,u \in V(G)$ where $u$ and $v$ are twins.
Consider the most permissive detector set, $S = V(G)$.
Because $u$ and $v$ are twins, either $N(u) = N(v)$, or $N[u] = N[v]$.
If $N(u) = N(v)$, then $uv \notin E(G)$ and we see that $N(u) \triangle N(v) = (N(u) \triangle N(v)) - \{u,v\} = \varnothing$.
Otherwise, $N[u] = N[v]$, so $uv \in E(G)$ and we find that $N[u] \triangle N[v] = (N(u) \triangle N(v)) - \{u,v\} = \varnothing$.
In either case, we see that property~\toroman{2} is violated.
Thus, $S = V(G)$ is not an ERR:LD set, so no ERR:LD set exists, completing the proof.
\end{proof}

\section{NP-completeness}\label{sec:np-complete}

Finding the minimum size of many detection system related parameters, such as OLD and IC sets, in arbitrary graphs are known to be NP-complete \cite{NP-complete-ic, old}.
It has also been proven that finding the minimum density of LD \cite{NP-complete-ld}, RED:LD \cite{redld}, and DET:LD \cite{detld} sets is similarly NP-complete.
We will now show that finding the smallest ERR:LD set is also NP-complete.

\npcompleteproblem{3-SAT}{Let $X$ be a set of $N$ variables.
Let $\psi$ be a conjunction of $M$ clauses, where each clause is a disjunction of three literals from distinct variables of $X$.}{Is there is an assignment of values to $X$ such that $\psi$ is true?}

\npcompleteproblem{Error-Correcting Locating-Domination (ERR-LD)}{A graph $G$ and integer $K$ with $2 \le K \le |V(G)|$.}{Is there exists an ERR:LD set $S$ with $|S| \le K$? Or equivalently, is ERR:LD($G$) $\le K$?}

\begin{theorem}
The ERR-LD problem is NP-complete.
\end{theorem}
\begin{wrapfigure}[13]{r}{0.37\textwidth}
    \centering
    \includegraphics[width=0.35\textwidth]{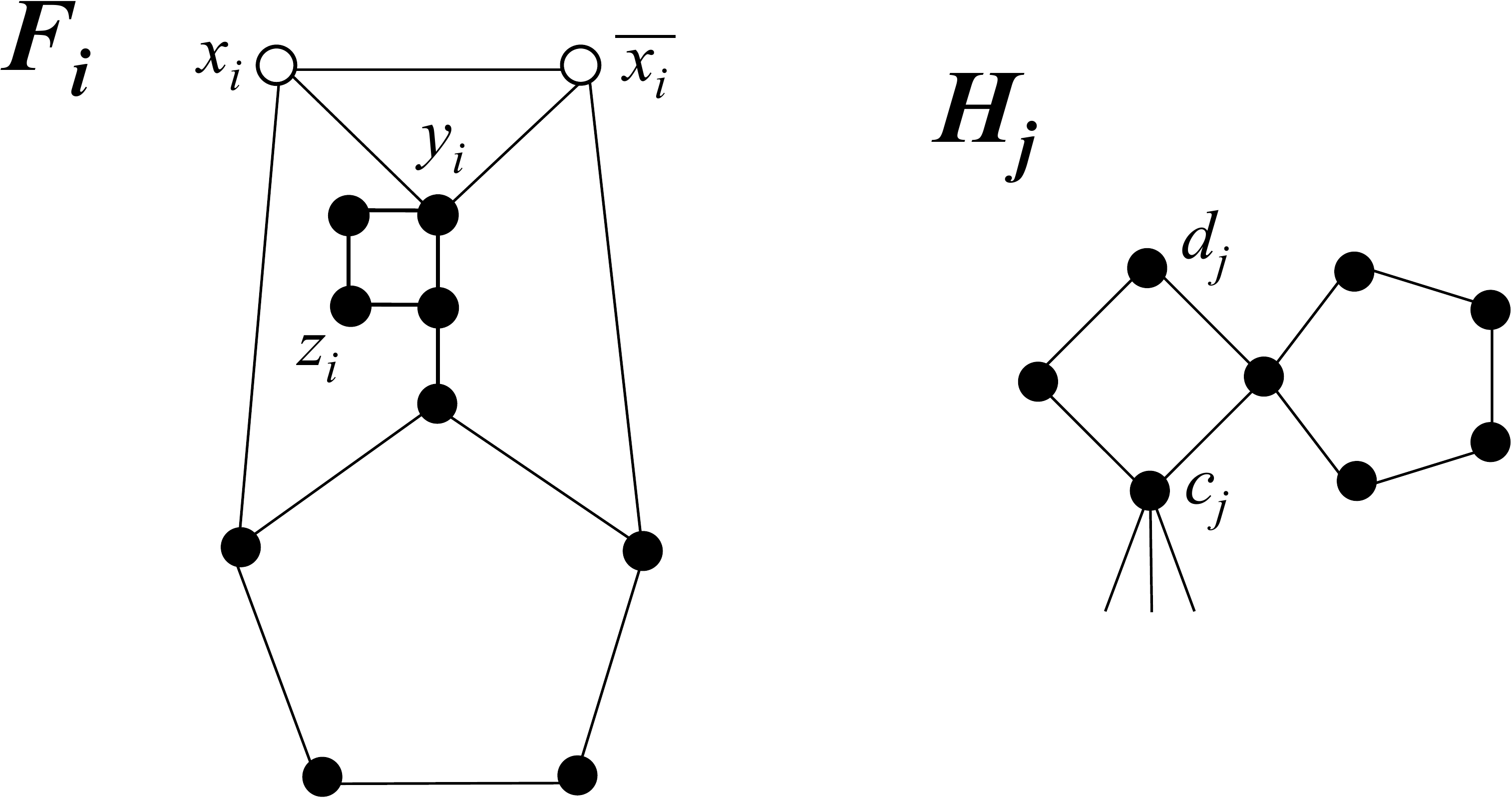}
    \caption{Variable and Clause graphs}
    \label{fig:variable-clause}
\end{wrapfigure}
\cbeginproof
Clearly, ERR-LD is NP, as every possible candidate solution can be generated nondeterministically in polynomial time (specifically, $O(n)$ time), and each candidate can be verified in polynomial time using Theorem~\ref{theo:err-ld-char}.
To complete the proof, we will now show a reduction from 3-SAT to ERR-LD.

\begin{figure}[ht]
    \centering
    \includegraphics[width=0.8\textwidth]{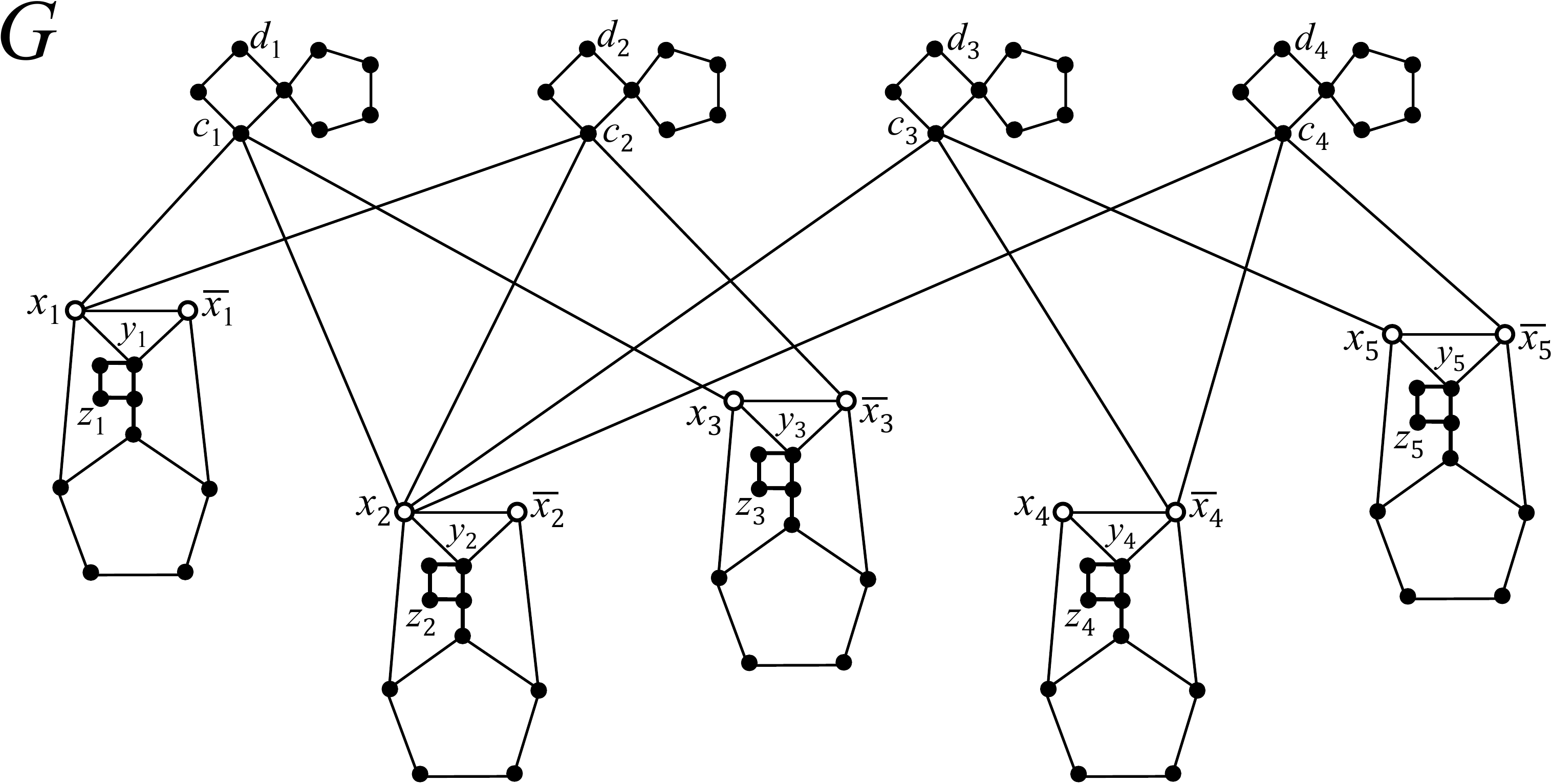}
    \caption{Construction of $G$ from \\ $(x_1 \lor x_2 \lor x_3) \land (x_1 \lor x_2 \lor \overline{x_3}) \land (x_2 \lor \overline{x_4} \lor x_5) \land (x_2 \lor \overline{x_4} \lor \overline{x_5})$}
    \label{fig:example-clause}
\end{figure}

Let $\psi$ be an instance of the 3-SAT problem with $M$ clauses on $N$ variables.
We will construct a graph, $G$, as follows.
For each variable $x_i$, create an instance of the $F_i$ graph (Figure~\ref{fig:variable-clause}); this includes a vertex for $x_i$ and its negation $\overline{x_i}$.
For each clause $c_j$ of $\psi$, create a new instance of the $H_j$ graph (Figure~\ref{fig:variable-clause}).
For each clause $c_j = \alpha \lor \beta \lor \gamma$, create an edge from the $c_j$ vertex to $\alpha$, $\beta$, and $\gamma$ from the variable graphs, each of which is either some $x_i$ or $\overline{x_i}$; for an example, see Figure~\ref{fig:example-clause}.
The resulting graph has precisely $11N + 8M$ vertices and $15N + 12M$ edges, and can be constructed in polynomial time.

Suppose $S \subseteq V(G)$ is an optimal (minimum) ERR:LD set on $G$.
By Theorem~\ref{theo:err-ld-char}, every vertex must be 3-dominated; thus, we require $9N + 8M$ detectors, as shown by the shaded vertices in Figure~\ref{fig:variable-clause}.
For any $F_i$, we require $\{x_i,\overline{x_i}\} \cap S \neq \varnothing$ to distinguish $y_i$ from $z_i$; similarly, we see that $c_j$ is not currently distinguished from $d_j$.
If $\{x_i,\overline{x_i}\} \cap S \neq \varnothing$ then we find that all vertices are 3-dominated and distinguished, completing the ERR:LD set.
Thus, we find that $|S| \ge 10N + 8M$; if $|S| = 10N + 8M$, then $|\{x_i,\overline{x_i}\} \cap S| = 1$ and $c_j$ must be dominated by one of its three neighbors in the $F_i$ graphs, so $\psi$ is satisfiable.

Now suppose we have a solution to the 3-SAT problem.
For each variable, $x_i$, if $x_i$ is true then we let the vertex $x_i \in S$; otherwise, we let $\overline{x_i} \in S$.
By construction, this 3-dominates and distinguishes every vertex in $G$; thus $S$ is an optimal ERR:LD set on $G$ of size $10N + 8M$.
\cendproof

\FloatBarrier
\section{ERR:LD sets in Special classes of graphs}

In this section, we explore ERR:LD sets in special classes of graphs.  
For a graph $G$, we let \textit{ERR:LD\%(G)} denote the minimum possible value of the density of a ERR:LD set, which is measured by the percentage of the vertices in the ERR:LD set over the entire vertices in $G$.
Formally, this is defined as $\limsup_{r \rightarrow \infty} \frac{|B_r(v) \cap S|}{|B_r(v)|}$ for any vertex $v \in V(G)$, where $B_r(v)$ is the ball of radius $r$ around $v$, $\{u \in V(G) : d(u,v) \le r\}$.

From Theorem~\ref{theo:err-ld-exist}, we see that graphs with a terminal vertex, for instance finite trees, cannot have an ERR:LD set because $\delta(G) \le 1$.
An ERR:LD set on the infinite path $P_\infty$ exists, but it requires every vertex be a detector due to the 3-dominating requirement of Theorem~\ref{theo:err-ld-char}.
Similarly, a cycle graph, $C_n$, has an ERR:LD set (requiring every vertex), but only exists for $n \ge 5$ since $C_3$ and $C_4$ contain twins.

\begin{observation}\label{theo:path-cycle}
If $G$ is 2-regular of order $n \ge 5$, like $P_\infty$ or $C_n$, then $\textrm{ERR:LD\%}(G) = 1$.
\end{observation}

\newpage
\unlabeledsection{ERR:LD sets in Cubic Graphs}

\textit{Cubic}, or 3-regular graphs, are graphs where each vertex has degree 3. 
Extensive work has been done on generation of cubic graphs \cite{bgm, bccs} and various domination-related  \cite{tot-dom-cubic} and detection system-related parameters \cite{loc-tot-dom-cubic, old-cubic} for these graphs.
For instance, Foucaud and Henning \cite{ld-cubic} proved that \emph{twin-free} cubic graphs have $\textrm{LD\%}(G) \le \frac{1}{2}$, and Seo \cite{ft-old-cubic} explored bounds for error-correcting open-locating-dominating (ERR:OLD) sets in cubic graphs.

First we establish the lower bound on ERR:LD in cubic graphs.

\begin{theorem}\label{theo:cubic-lower}
If $G$ is a twin-free cubic graph, then $\textrm{ERR:LD\%}(G) \ge \frac{3}{4}$.
\end{theorem}
\begin{proof}
Because $G$ is 3-regular, $|N[v]| = 4$ for any vertex $v$, and an ERR:LD set must 3-dominate each vertex.
Thus, we need at least three of the four vertices around any vertex to be detectors, completing the proof.
\end{proof}

\begin{figure}[ht]
    \centering
    \includegraphics[width=0.8\textwidth]{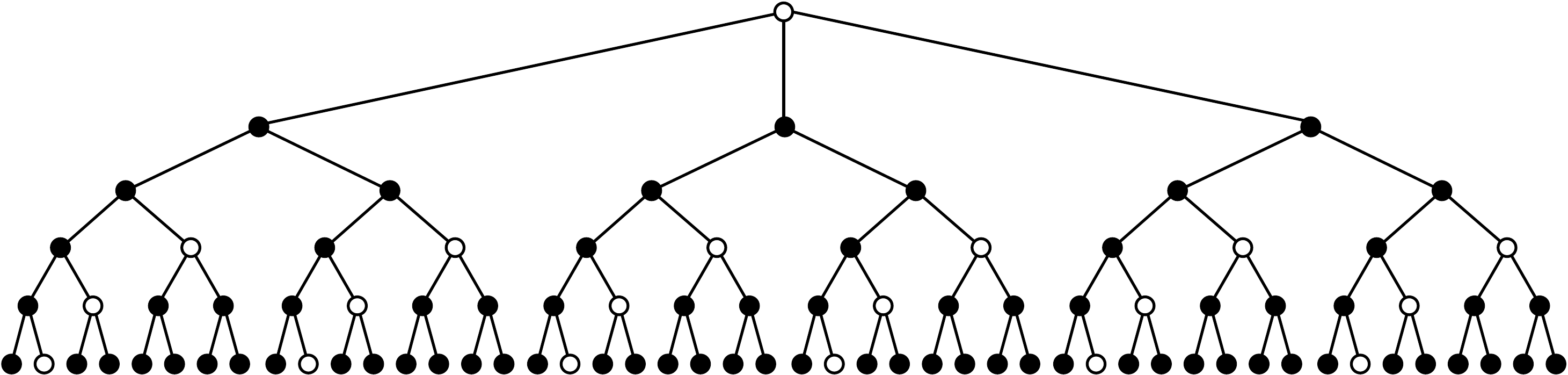}
    \caption{Optimal DET:LD set on the infinite 3-regular tree with density $\frac{3}{4}$.}
    \label{fig:3-reg-tree-soln}
\end{figure}

\begin{theorem}
The infinite 3-regular tree has $\textrm{ERR:LD\%}(G) = \frac{3}{4}$.
\end{theorem}
\cbeginproof
From Theorem~\ref{theo:cubic-lower}, we have that $\textrm{ERR:LD\%}(G) \ge \frac{3}{4}$, so we need only show that $\textrm{ERR:LD\%}(G) \le \frac{3}{4}$.
We will now construct a subset of vertices, $S \subseteq V(G)$.
Let $r \in V(G)$ be some starting vertex in the tree.
Let $r \notin S$ and $N(r) \subseteq S$, and mark all vertices in $N[r]$ as ``visited".
Then, for any visited vertex $v$ which has one or more unvisited neighbors, add detectors at unvisited vertices of $N(v)$ to cause $|N[v] \cap S| = 3$, and mark all vertices in $N[v]$ as visited.
From this construction, it is clear that every vertex is exactly 3-dominated, giving us a density of $\frac{3}{4}$.
We also see that every pair of vertices is distinguished, so $S$ is an ERR:LD set.
Therefore, $\textrm{ERR:LD\%}(G) \le \frac{3}{4}$, completing the proof.
\cendproof

\begin{figure}[ht]
    \centering
    \includegraphics[width=0.45\textwidth]{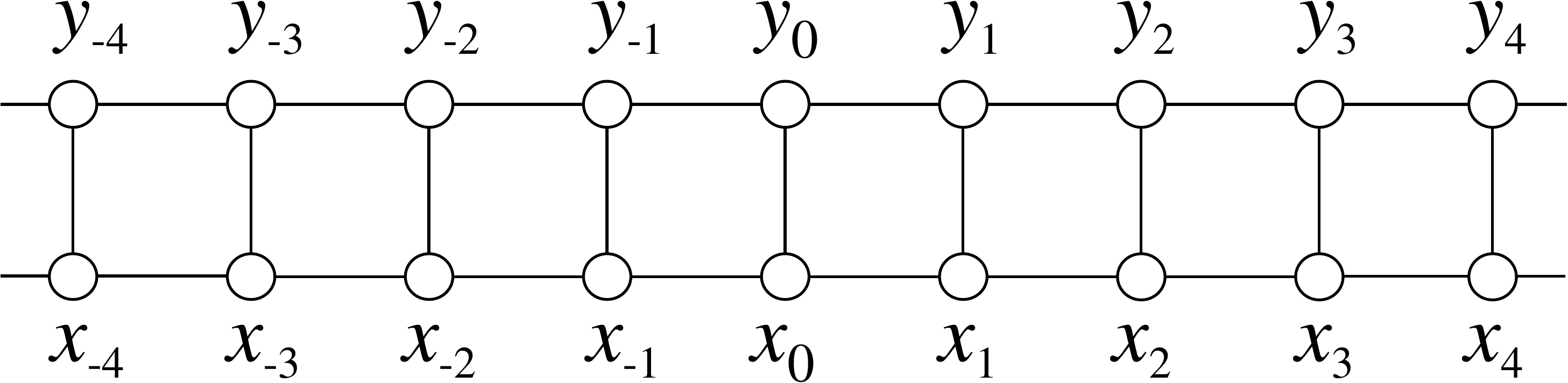}
    \caption{Ladder graph labeling scheme}
    \label{fig:ladder-labeling}
\end{figure}

\FloatBarrier
\newpage
\begin{theorem}
The infinite ladder graph has $\textrm{ERR:LD\%}(P_\infty \square P_2) = \frac{5}{6}$.
\end{theorem}
\cbeginproof
Refer to the labeling scheme given by Figure~\ref{fig:ladder-labeling}.
To prove a lower bound of $\frac{5}{6}$, we will take any non-detector vertex, $v \notin S$, and uniquely associate five detectors with it.
For the association, we say that $v$ can be associated with any vertex in $\{u \in V(G) : u \in N(v) \textrm{ or } |N(u) \cap N(v)| = 2\}$.
To begin, suppose $y_0 \notin S$ is a non-detector vertex; to 3-dominate $y_0$ we need $\{y_{-1},y_1,x_0\} \subseteq S$.
To 3-dominate $y_1$ we require $\{y_2,x_1\} \subseteq S$, and by symmetry $\{y_{-2},x_{-1}\} \subseteq S$ as well.
To distinguish $y_0$ and $x_1$, we require $x_2 \in S$, and by symmetry $x_{-2} \in S$.
From the association scheme described above, we see that $y_0$ can be associated with five detectors, $\{y_{-1},y_1,x_{-1},x_0,x_1\}$, and this association is unique because the closest non-detector to $y_0$ is potentially $y_3$, $y_{-3}$, $x_3$, or $x_{-3}$ and none can be associated with any of these five vertices.
Thus, $\textrm{ERR:LD}(P_\infty \square P_2) \ge \frac{5}{6}$.

\begin{figure}[ht]
    \centering
    \includegraphics[width=0.6\textwidth]{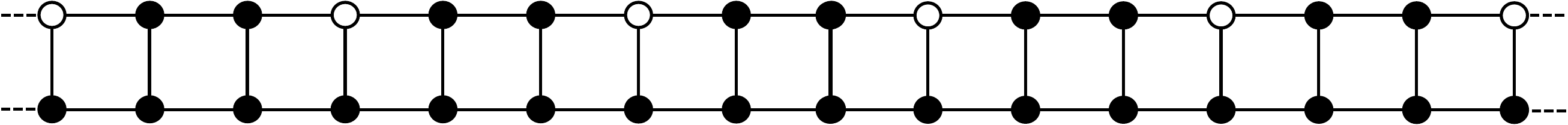}
    \caption{An optimal ERR:LD set on the infinite ladder graph with density $\frac{5}{6}$}
    \label{fig:ladder-soln}
\end{figure}

For the upper bound, let $S = \{x_i : i \in \mathbb{Z}\} \cup \{y_i : i \mod 3 = 0\}$; this construction is given by Figure~\ref{fig:ladder-soln}.
Then every vertex 3-dominated, and each vertex pair is distinguished as per Theorem~\ref{theo:err-ld-char}, so $S$ is an ERR:LD set.
Additionally, we see that $S$ has a density of $\frac{5}{6}$, so $\textrm{ERR:LD}(P_\infty \square P_2) \le \frac{5}{6}$, completing the proof.
\cendproof

\begin{theorem}\label{theo:dist-5}
Let $G$ be a twin-free cubic graph.
If $\overline{S} \subseteq V(G)$ with $\forall x,y \in \overline{S}$, $d(x,y) \ge 5$, then $S = V(G) - \overline{S}$ is a ERR:LD set for $G$.
\end{theorem}
\begin{proof}
Firstly, note that for any $w \in V(G)$, there can be at most one non-detector in $N[w]$ because they are all within distance $5$ of one another, so $|N[w] \cap S| \ge 3$.
Let $u, v \in V(G)$.
Suppose $d(u, v) \le 2$; then because $G$ is twin-free and regular, $\exists w_1 \in N(u) - N[v]$ and $\exists w_2 \in N(v) - N[u]$.
Because $d(u,v) \le 2$, $\{u,v\} \cap S \neq \varnothing$.
If $u \notin S$ or by symmetry $v \notin S$, then $\{w_1,w_2\} \subseteq S$, so $u$ and $v$ are 2-distinguished.
Otherwise, $\{u,v\} \subseteq S$, and because they are all within distance 5, $\{w_1,w_2\} \cap S \neq \varnothing$, so $u$ and $v$ are 1-distinguished.
Otherwise $d(u, v) \ge 3$, so $N(u) \triangle N(v) = \varnothing$.
At most one detector in each of $N[v]$ and $N[u]$ can be a non-detector, so $u$ and $v$ are at least 4-distinguished.
Thus, in any case, $u$ and $v$ are both 3-dominated and distinguished, so $S = V(G) - \overline{S}$ is an ERR:LD set for $G$.
\end{proof}

For a cubic graph $G$ and a vertex $v \in V(G)$, we observe that there are at most 3, 6, 12, and 24 vertices at distance 1, 2, 3, and 4 from $v$, respectively.
Therefore, there are at most 45 vertices at distance less than 5, and hence by using Theorem \ref{theo:dist-5} we have a simple upper bound for ERR:LD($G$) for a cubic graph $G$ as shown in the next proposition.

\begin{prop}
If $G$ is a twin-free cubic graph of order $n \geq 46$, then  $ERR$:$LD(G) \leq \ceil{\frac{45}{46}n}$.
\end{prop}

\newpage
\unlabeledsection{ERR:LD sets in the infinite grids}

In this section, we will look at ERR:LD sets on four of the most studied infinite grids, namely the square, hexagonal, triangular, and king grids.
The bibliography maintained by Lobstein \cite{dombib} contains numerous articles on various detection-system-related parameters in these infinite grids.

\begin{figure}[ht]
    \centering
    \begin{tabular}{cc}
        \includegraphics[width=0.4\textwidth]{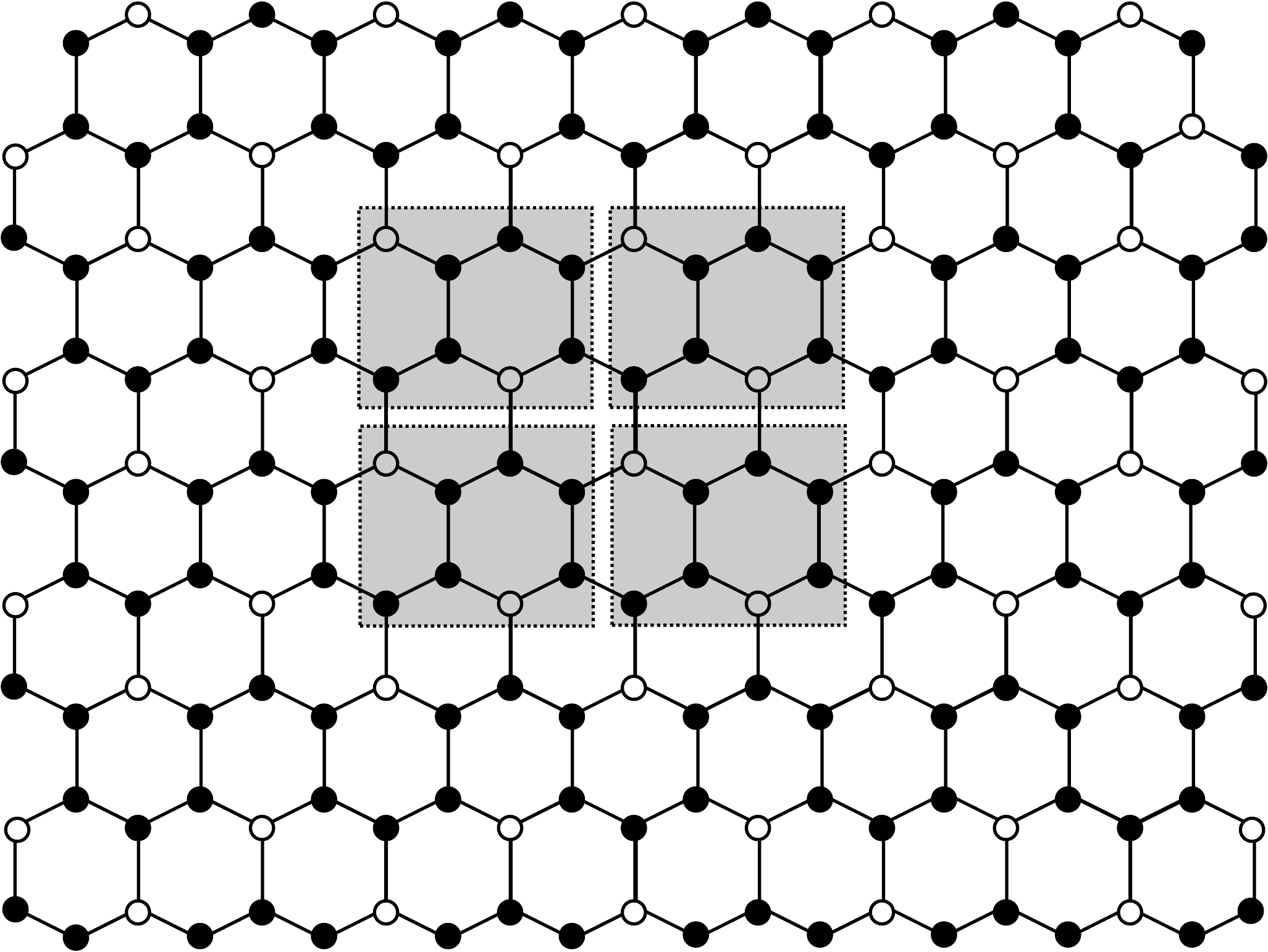} & \includegraphics[width=0.4\textwidth]{fig/sq-err-ld.pdf} \\ (a) & (b) \\ \\
        \includegraphics[width=0.455\textwidth]{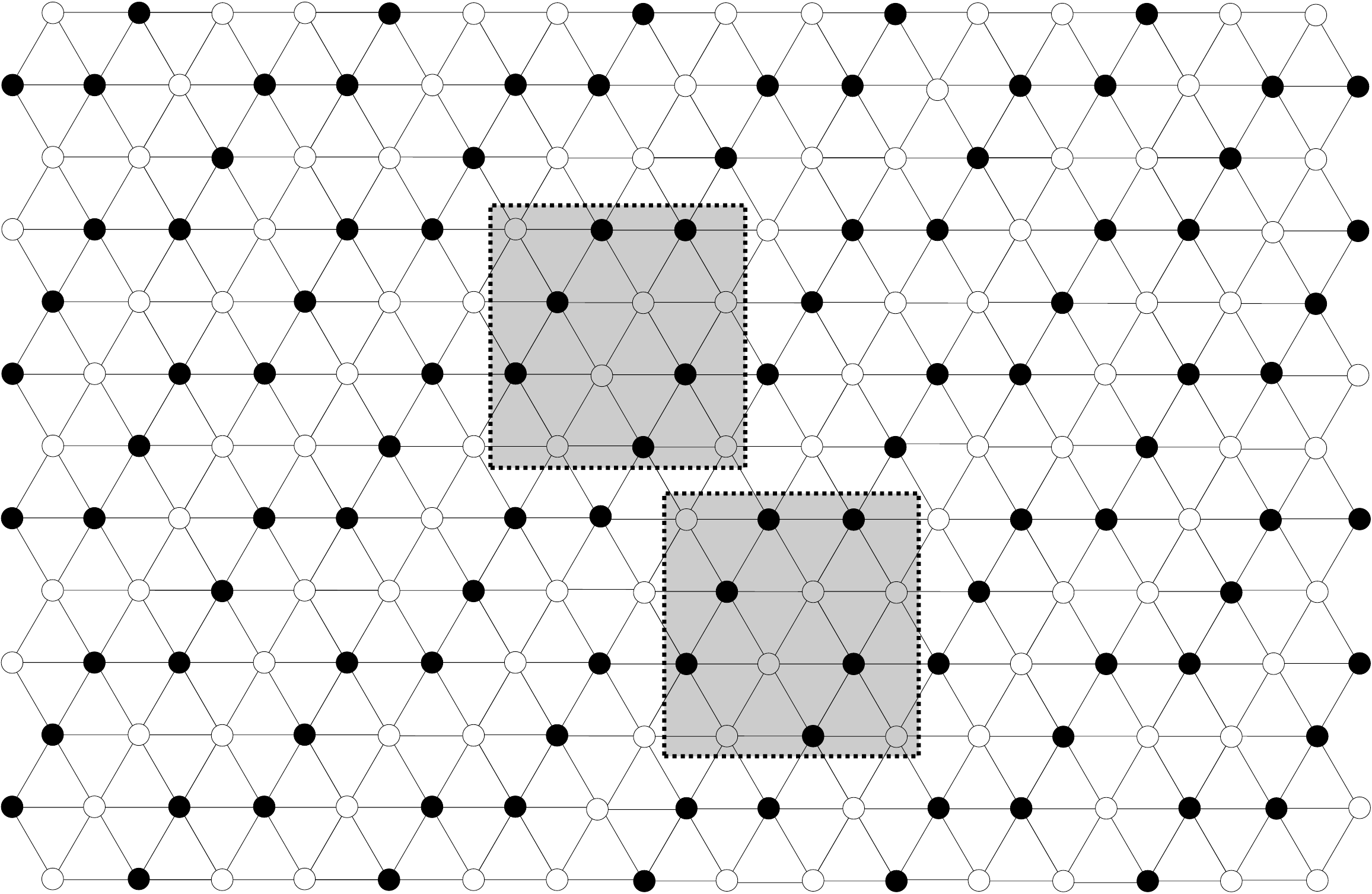} &  \includegraphics[width=0.4\textwidth]{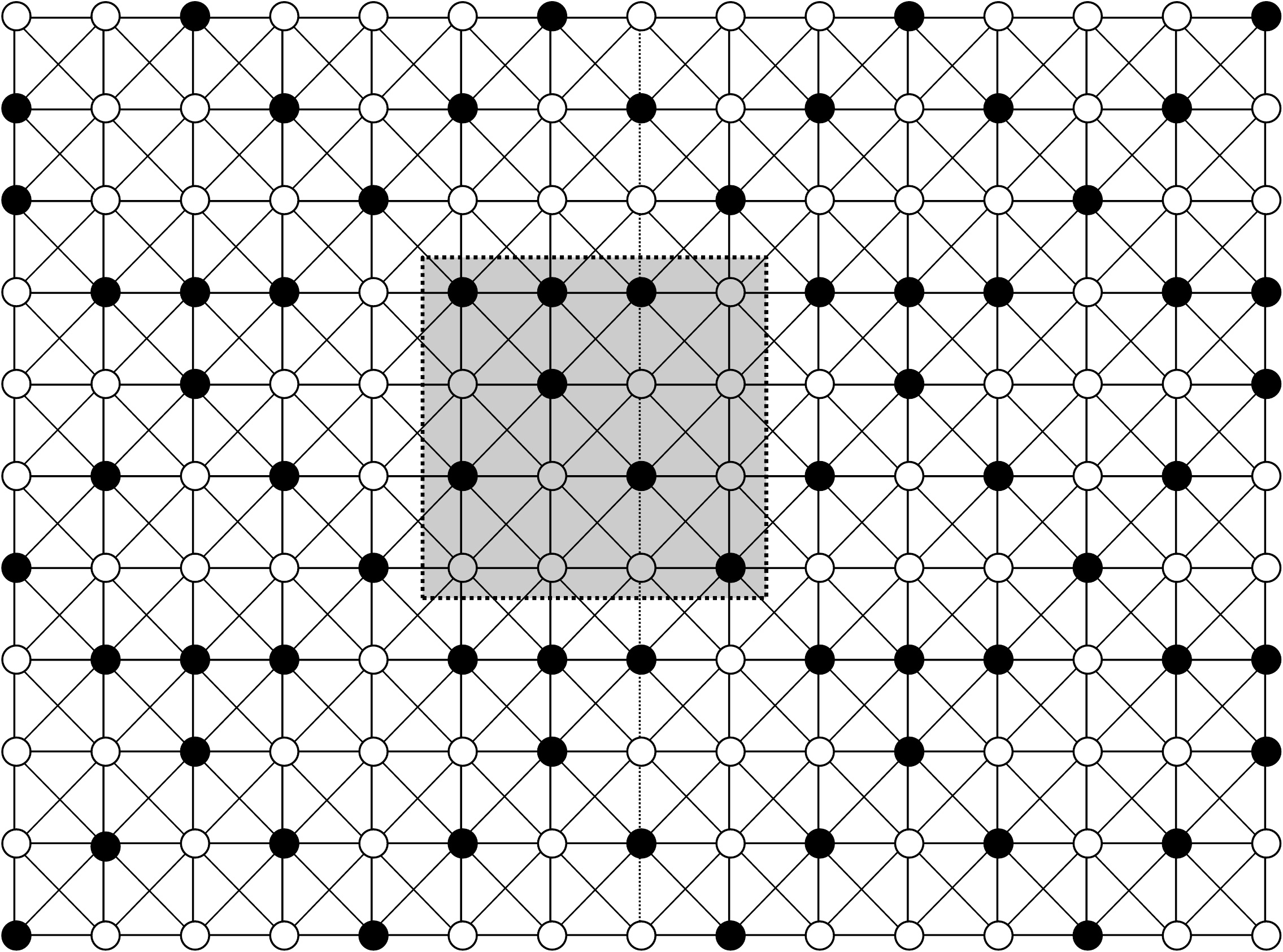} \\ (c) & (d)
    \end{tabular}
    \caption{Upper bounds for the ERR:LD on (a) $HEX$, (b) $TRI$, (c) $SQ$, and (d) $K$.}
    \label{fig:inf-grids-upper-bounds}
\end{figure}

Previous work on detection system related parameters in the \emph{infinite hexagonal grid}, HEX, include Honkala and Laihonen \cite{ld-hex-king}, who proved that LD\%(HEX) $= \frac{1}{3}$.
We \cite{redld, detld} have shown that RED:LD\%(HEX) $= \frac{1}{2}$ and $\frac{3}{5} < \textrm{DET:LD\%}(\textrm{HEX}) \le \frac{5}{8}$.
Figure \ref{fig:inf-grids-upper-bounds}~(a) shows an ERR:LD set for HEX with density $\frac{3}{4}$, so we have $\textrm{ERR:LD\%}(\textrm{HEX}) \le \frac{3}{4}$.
Since HEX is cubic, Theorem~\ref{theo:cubic-lower} from above gives us $\textrm{ERR:LD\%}(\textrm{HEX}) \ge \frac{3}{4}$.

\begin{theorem}
$\textrm{ERR:LD}\%(HEX) = \frac{3}{4}$
\end{theorem}

Recall that SQ is the infinite square grid, a 4-regular graph.
Slater \cite{ftld} previously determined that $\textrm{LD}(SQ) = \frac{3}{10}$, and we \cite{redld} have shown $\frac{2}{5} \le \textrm{RED:LD}(SQ) \le \frac{7}{16}$.
Figure \ref{fig:inf-grids-upper-bounds}~(b) shows an ERR:LD set for the SQ graph with density $\frac{2}{3}$, so we have ERR:LD\%(SQ) $\le \frac{2}{3}$. 
By Theorem~\ref{theo:err-ld-char}, every vertex in an ERR:LD set must be at least 3-dominated; thus, $\forall v \in V(G)$, $|N[v] \cap S| \ge 3$, meaning at least three of the five vertices in $N[v]$ must be detectors, resulting in a lower bound of $\frac{3}{5}$.
We have proven a better lower bound of $\frac{24}{37}$, but have elected not to include its proof due to the length.

\begin{theorem}
$\frac{24}{37} \leq \textrm{ERR:LD}\%(SQ) \leq \frac{3}{4}$
\end{theorem}

The infinite triangular grid, denoted as TRI, is a 6-regular graph. 
Honkala \cite{ldtri} determined $\textrm{LD}(TRI) = \frac{13}{57}$, and we \cite{redld} have shown $\frac{2}{5} \le \textrm{RED:LD}(TRI) \le \frac{7}{16}$.
Figure \ref{fig:inf-grids-upper-bounds}~(c) shows an ERR:LD set for the TRI graph with density $\frac{1}{2}$, so we have ERR:LD\%(TRI) $\le \frac{1}{2}$. 
By Theorem~\ref{theo:err-ld-char}, every vertex in an ERR:LD set must be at least 3-dominated; thus, at least three of the seven vertices in $N[v]$ must be detectors, resulting in a lower bound of $\frac{3}{7}$.

\begin{theorem}
$\frac{3}{7} \leq \textrm{ERR:LD}\%(TRI) \leq \frac{1}{2}$
\end{theorem}

Now we consider the \emph{infinite king grid}, K, which is an 8-regular graph based on the legal movements of a king on a chess board.
Honkala and Laihonen \cite{ld-hex-king} proved $\textrm{LD}(K) = \frac{1}{5}$, and Jean and Seo \cite{redld} established the bounds for $K$: $\frac{3}{11} \le \textrm{RED:LD}(K) \le \frac{5}{16}$.
Figure \ref{fig:inf-grids-upper-bounds}~(d) shows an ERR:LD set for the king graph with density $\frac{7}{16}$, so we have ERR:LD\%(K) $\le \frac{7}{16}$. 
By Theorem~\ref{theo:err-ld-char}, every vertex in an ERR:LD set must be at least 3-dominated; thus, at least three of the nine vertices in $N[v]$ must be detectors, resulting in a lower bound of $\frac{1}{3}$.

\begin{theorem}
$\frac{1}{3} \leq \textrm{ERR:LD}\%(K) \leq \frac{7}{16}$
\end{theorem}

\FloatBarrier
\vspace{0em}
\section{Summary and Future Directions}\label{sec:future}

In this paper, we have investigated a fault-tolerant detection system modeled as an error-correcting locating-dominating set.
After introducing the parameter, we proved the problem of determining the value of ERR:LD(G) for an arbitrary graph is NP-complete. 
Next, we fully characterized the parameter and established criteria for the existence of such a set.
The existence criteria enable us to determine whether an ERR:LD exists for a given graph, $G$; the characterization is useful in finding ERR:LD sets for $G$ and determining the value of ERR:LD($G$).

Further, we explored some special classes of graphs, including cubic graphs and four infinite grid graphs to find optimal (minimum) ERR:LD sets and their densities.
For cubic graphs, we established the lower and upper bounds of the value of ERR:LD and proved the lower bound is sharp.  
We also establish bounds for $\textrm{ERR:LD}(G)$ for all four examined infinite grids, including a tight bound for the hexagonal grid.

We believe that we can improve the upper bound of ERR:LD established for cubic graphs, and the problem is under investigation.
Characterizing cubic graphs that achieve the lower and upper bounds on ERR:LD is also an interesting problem that is under study.
Another possibility is to improve the lower bounds we have presented for ERR:LD\%(TRI) and ERR:LD\%(K).
We have not found the tight bound for the square grid, but the lower bound we have obtained is very close to the upper bound.
To date, there is no proof of ERR:LD\%(SQ) $= \frac{2}{3}$, but our further examination strongly suggests that $\frac{2}{3}$ is the tight bound.
Therefore, we have the following conjecture, and we are working on the proof.

\begin{conjecture}
$\textrm{ERR:LD}(SQ) = \frac{2}{3}$.
\end{conjecture}

\vspace{0em}
\section{Proofs of Characterization}\label{sec:proofs}

In the following proofs for characterizing ERR:LD sets, we say that each detector "transmits" a value to indicate the location of an intruder it detects, if any.
We have arbitrarily elected for transmitting 0 for no intruder being detected by the sensor, 1 for an intruder in $N(v)$, or 2 for an intruder at its location, $\{v\}$.
For organization, we will first prove two lemmas which together give the necessary conditions for a ERR:LD set, and then finalize the characterization by proving these conditions are sufficient in Theorem~\ref{theo:err-ld-char}.

\begin{lemma}\label{lem:err-ld-s-imp}
Let $S \subseteq V(G)$ be an ERR:LD set and $v \in S$. Then
\begin{enumerate}[noitemsep, label=\roman*.]
    \item $|N(v) \cap S| \ge 2$
    \item $\forall u \in S$ with $u \neq v$, $|((N(v) \cap S) \triangle (N(u) \cap S)) - \{v, u\}| \ge 1$
\end{enumerate}
\end{lemma}
\begin{proof}
Suppose that property~\toroman{1} is false; then $\exists v \in S$ such that $|N(v) \cap S| \le 1$.
By definition, an ERR:LD set is a DET:LD set, so Theorem~\ref{theo:det-ld-char} yields that $|N(v) \cap S| \ge 1$; thus $|N(v) \cap S| = 1$, meaning $N(v) \cap S = \{u\}$ for some $u \in S$ with $u \neq v$.
Suppose there is not an intruder but $v$ incorrectly transmits 2; then
the system will not be able to distinguish this scenario from there being an intruder at $v$ and $u$ incorrectly transmitting 0.
Thus, the intruder is not found, a contradiction.

Suppose that property~\toroman{2} is false; then $\exists u \in S$ with $u \neq v$ such that $((N(v) \cap S) \triangle (N(u) \cap S)) - \{v, u\} = \varnothing$.
Theorem~\ref{theo:det-ld-char} yields that $|(N(v) \cap S) \triangle (N(u) \cap S)| \ge 1$, so $((N(v) \cap S) \triangle (N(u) \cap S)) \subseteq \{v,u\}$, meaning $u \in N(v)$ and $v \in N(u)$.
Thus, there are no detectors in the open neighborhoods of $v$ and $u$ other than $v$ and $u$ themselves.
Suppose there is an intruder at $v$, and $v$ incorrectly transmits 1; then the system will not be able to distinguish this from an intruder being at $u$ and $u$ incorrectly transmitting 1.
Thus, the intruder is not found, a contradiction.
\end{proof}

\begin{lemma}\label{lem:err-ld-not-s-imp}
Let $S \subseteq V(G)$ be an ERR:LD set and $v \notin S$. Then
\begin{enumerate}[noitemsep, label=\roman*.]
    \item $|N(v) \cap S| \ge 3$
    \item $\forall u \in S$, $|((N(v) \cap S) \triangle (N(u) \cap S)) - \{u\}| \ge 2$
    \item $\forall u \in V(G)-S$ with $u \neq v$, $|(N(v) \cap S) \triangle (N(u) \cap S)| \ge 3$
\end{enumerate}
\end{lemma}
\begin{proof}
Suppose property~\toroman{1} is false; then $\exists v \in V(G)-S$ with $|N(v) \cap S| \le 2$.
Theorem~\ref{theo:det-ld-char} yields that $|N(v) \cap S| \ge 2$, so $|N(v) \cap S| = 2$.
Thus, $N(v) \cap S = \{x,y\}$ for some $x,y \in S$ with $x \neq y$.
Suppose there is no intruder but $x$ incorrectly transmits 1; then the system will not be able to distinguish this from there being an intruder at $v$ and $y$ incorrectly transmitting 0.
Thus, the intruder is not found, a contradiction.

We will now prove property~\toroman{2}; let $v \in V(G)-S$ and $u \in S$.
By definition, every ERR:LD set is a DET:LD set, and by Corollary~\ref{cor:det-is-red} is a RED:LD set; thus, Theorem~\ref{theo:red-ld-char} yields that $|((N(v) \cap S) \triangle (N(u) \cap S)) - \{u\}| \ge 1$.
Suppose $|((N(v) \cap S) \triangle (N(u) \cap S)) - \{u\}| = 1$; then $\exists t \in ((N(v) \cap S) \triangle (N(u) \cap S)) - \{u\}$ with $t \in N(v)$ or $t \in N(u)$, but not both.
Consider when $t \in N(v)$ and $u \notin N(v)$; if there is an intruder at $u$ but $u$ incorrectly transmits 0, then the system cannot distinguish this from the intruder being at $v$ and $t$ incorrectly transmitting 0; thus, the intruder is not found, a contradiction.
Consider when $t \in N(v)$ and $u \in N(v)$; if there is an intruder at $u$ but $u$ incorrectly transmits 1, then the system cannot distinguish this from the intruder being at $v$ and $t$ incorrectly transmitting 0; thus, the intruder is not found, a contradiction.
Lastly, consider when $t \in N(u)$; regardless of whether $u \in N(v)$, if there is an intruder at $v$ and $u$ incorrectly transmits 2, then the system cannot distinguish this from an intruder being at $u$ and $t$ incorrectly transmitting 0.
Thus, the intruder is not found, a contradiction.
In any case we have a contradiction; therefore, $|((N(v) \cap S) \triangle (N(u) \cap S)) - \{u\}| \ge 2$.

Suppose property~\toroman{3} is false; then $\exists v,u \in V(G)-S$ with $u \neq v$ such that $|(N(v) \cap S) \triangle (N(u) \cap S)| \le 2$.
Theorem~\ref{theo:det-ld-char} yields that $|(N(v) \cap S) - (N(u) \cap S)| \ge 2$ or $|(N(u) \cap S) - (N(v) \cap S)| \ge 2$.
As the symmetric difference is the union of these two (disjoint) differences, it must be the case that one difference is of size 2 and the other is of size 0; without loss of generality, let $|(N(v) \cap S) - (N(u) \cap S)| = 0$ and $|(N(u) \cap S) - (N(v) \cap S)| = 2$, so $(N(u) \cap S) - (N(v) \cap S) = \{x,y\}$ for some $x,y \in S$ with $x \neq y$.
If there is an intruder at $v$ and $x$ incorrectly transmits 1, then the system cannot determine whether there is an intruder at $v$ and $x$ incorrectly transmitted 1 or the intruder is at $u$ and $y$ incorrectly transmitted 0; thus, the intruder is not found, a contradiction.
Therefore, $|(N(v) \cap S) \triangle (N(u) \cap S)| \ge 3$.
\end{proof}

\begin{theorem} 
A detector set $S \subseteq V(G)$ is an ERR:LD set if and only if the following are true:
\begin{enumerate}[noitemsep, label=\roman*.]
    \item $\forall v \in V(G)$, $|N[v] \cap S| \ge 3$
    \item $\forall v,u \in S$ with $u \neq v$, $|((N(v) \cap S) \triangle (N(u) \cap S)) - \{v, u\}| \ge 1$
    \item $\forall v \in V(G)-S$ and $\forall u \in S$, $|((N(v) \cap S) \triangle (N(u) \cap S)) - \{u\}| \ge 2$
    \item $\forall v,u \in V(G)-S$ with $u \neq v$, $|(N(v) \cap S) \triangle (N(u) \cap S)| \ge 3$
\end{enumerate}
\end{theorem}
\begin{proof}
Lemmas \ref{lem:err-ld-s-imp} and \ref{lem:err-ld-not-s-imp} prove that every ERR:LD set satisfies properties \toroman{1}--\toroman{4}.
For the converse, suppose detector set $S \subseteq V(G)$ satisfies properties \toroman{1}--\toroman{4}.
We now consider any possible output from the detectors; because we assume there is at most one error and one intruder, it is impossible to have more than two detectors transmit 2.
Therefore, we consider cases where 0, 1, or 2 detectors transmit 2.
Let $A = \{w \in S : w \textrm{ transmits } 1\}$.

Suppose no detector transmits 2.
By property~\toroman{1}, every vertex is at least 3-dominated, and by hypothesis there can be at most one false negative.
Thus, if $|A| \le 1$, we determine that there is no intruder; otherwise, $|A| \ge 2$ and we know that an intruder is present.
Let $p,q \in V(G)$ with $p \neq q$.
If $|(N[p] \cap S) - A| \ge 2$, then the intruder is not at $p$.
Similarly, if $|(N[q] \cap S) - A| \ge 2$, then the intruder is not at $q$.
Otherwise, we assume $|(N[p] \cap S) - A| \le 1$ and $|(N[q] \cap S) - A| \le 1$.
\textbf{Case~1:} $p \in S$ and $q \in S$.
By property~\toroman{2}, $\exists w \in ((N(p) \cap S) \triangle (N(q) \cap S)) - \{p,q\}$; due to symmetry, without loss of generality let $w \in N(p)$.
If $w \notin A$ then $w$ transmits 0 and we know the intruder is not at $p$, as otherwise this would require $p$ incorrectly transmit 1 and $w$ incorrectly transmit 0, contradicting that there can be at most one error.
Otherwise $w \in A$, which means the intruder cannot be at $q$ without both $q$ and $w$ being faulty, a contradiction.
Thus, in any configuration we can always eliminate at least one of $p$ or $q$.
\textbf{Case~2:} $p \in S \oplus q \in S$; without loss of generality let $p \in S$ and $q \notin S$.
By property~\toroman{3}, $\exists x,y \in ((N(p) \cap S) \triangle (N(q) \cap S)) - \{p\}$ with $x \neq y$.
If $x \in N(p)$ and $x \notin A$ or if $y \in N(p)$ and $y \notin A$, then the intruder is not at $p$, as otherwise there would be more than one error; thus, we assume $\{x,y\} \cap N(p) \subseteq A$.
Similarly, if $x \in N(q)$ and $x \in A$ or if $y \in N(q)$ and $y \in A$, then the intruder is not at $p$; otherwise we assume $\{x,y\} \cap N(q) \cap A = \varnothing$.
If $\{x,y\} \subseteq N(p)$, then we observe the intruder cannot be at $q$ without both $x$ and $y$ being faulty; otherwise $|\{x,y\} \cap N(q)| \ge 1$ and we find that the intruder cannot be at $p$ without having two or more errors.
Thus, in any configuration we can always eliminate at least one of $p$ or $q$.
\textbf{Case~3:} $p \notin S$ and $q \notin S$.
By property~\toroman{4}, $\exists x,y,z \in (N(p) \cap S) \triangle (N(q) \cap S)$ with $x \neq y$, $y \neq z$, and $z \neq x$.
Without loss of generality, let $|\{x,y,z\} \cap N(p)| > |\{x,y,z\} \cap N(q)|$.
If $|\{x,y,z\} \cap N(p) \cap A| \ge 2$, then the intruder cannot be at $q$; otherwise we assume $|\{x,y,z\} \cap N(p) \cap A| \le 1$.
If $\{x,y,z\} \subseteq N(p)$, then $|\{x,y,z\} \cap A| \le 1$, implying $|\{x,y,z\} - A| \ge 2$; thus, we find that the intruder cannot be at $p$.
Otherwise $|\{x,y,z\} \cap N(p)| = 2$; without loss of generality let $\{x,y\} \subseteq N(p)$ and $z \in N(q)$.
Because $|\{x,y,z\} \cap N(p) \cap A| \le 1$, $|\{x,y\} \cap A| \le 1$; without loss of generality, let $x \notin A$.
If $y \notin A$, then the intruder cannot be at $p$ without both $x$ and $y$ being faulty; otherwise we assume $y \in A$.
If $z \in A$, then the intruder cannot be at $p$ without both $x$ and $z$ being faulty; otherwise $z \notin A$ and we see that the intruder cannot be at $q$ without both $y$ and $z$ being faulty, a contradiction.
Thus, in any configuration we see we can eliminate at least one of $p$ or $q$.

Suppose exactly one detector $v \in S$ transmits 2.
We assume there can be at most one error, and by property~\toroman{1} all vertices are at least 2-open-dominated.
Thus, if $|A| = 0$, we determine that $v$ incorrectly transmitted 2 and there is no intruder; otherwise, $|A| \ge 1$ and we know an intruder is present.
By property~\toroman{1}, all vertices are at least 3-dominated; thus, if $|A| = 1$, then $A \subseteq N(v) \cap S$ and we determine that the intruder is at $v$ and one of its neighbors incorrectly transmitted 0; otherwise we assume $|A| \ge 2$.
Let $p,q \in V(G)$ with $p \neq q$.
\textbf{Case~1:} $p \in S$ and $q \in S$.
If $v \notin N[p] \cup N[q]$, then the intruder cannot be at $p$ without both $p$ and $v$ being faulty; otherwise we assume $v \in N[p] \cup N[q]$.
If $p = v$, then the intruder cannot be at $q$ without both $p$ and $q$ being faulty; otherwise we assume $p \neq v$.
By symmetry, if $q = v$, then the intruder cannot be at $p$; otherwise we assume $q \neq v$.
By property~\toroman{2}, $\exists w \in ((N(p) \cap S) \triangle (N(q) \cap S)) - \{p,q\}$; without loss of generality, let $w \in N(p)$.
If $w$ transmits 0, then the intruder cannot be at $p$, as otherwise both $p$ and $w$ would be faulty, a contradiction.
Otherwise $w$ transmits 1 or 2; in either case the intruder cannot be at $q$ without both $q$ and $w$ being faulty, a contradiction.
Thus, in any configuration we can always eliminate at least one of $p$ or $q$.
\textbf{Case~2:} $p \in S \oplus q \in S$; without loss of generality, let $p \in S$ and $q \notin S$.
Suppose $v \notin N[p] \cup N[q]$; then $p$ cannot be the intruder location without both $p$ and $v$ being faulty; otherwise we assume $v \in N[p] \cup N[q]$.
By property~\toroman{3}, $\exists x,y \in ((N(p) \cap S) \triangle (N(q) \cap S)) - \{p\}$ with $x \neq y$.
Suppose $p = v$.
If $\{x,y\} \cap N(p) \cap A \neq \varnothing$, then the intruder cannot be at $q$ without there being more than one error, a contradiction; thus, we assume $\{x,y\} \cap N(p) \cap A = \varnothing$.
If $\{x,y\} \subseteq N(p)$, then the intruder cannot be at $p$, as otherwise both $x$ and $y$ would be faulty, a contradiction; thus, we assume $|\{x,y\} \cap N(p)| \le 1$, meaning $|\{x,y\} \cap N(q)| \ge 1$.
If $|\{x,y\} \cap N(p)| = 1$, then without loss of generality let $x \in N(p)$ and $y \in N(q)$; if $y \in A$ then the intruder cannot be at $p$ without both $x$ and $y$ being faulty; otherwise $y \notin A$ and we find that the intruder cannot be at $q$ without both $p$ and $y$ being faulty.
Otherwise we assume $|\{x,y\} \cap N(p)| = 0$, implying $\{x,y\} \subseteq N(q)$.
If $\{x,y\} \subseteq A$, then the intruder cannot be at $p$ without both $x$ and $y$ being faulty.
If $\{x,y\} \cap A = \varnothing$, then the intruder cannot be at $q$ without both $x$ and $y$ being faulty.
Otherwise $|\{x,y\} \cap A| = 1$, and without loss of generality let $x \in A$ and $y \notin A$; then the intruder cannot be at $q$ without both $p$ and $y$ being faulty.
Thus, in any configuration where $p = v$ we can always eliminate $p$ or $q$.
Now we consider the other possibility, where $p \neq v$.
Because we assume $v \in N[p] \cup N[q]$ and $p \neq v$, we know that $v \in (N(p) \cup N(q)) - \{p\}$.
We observe that the intruder cannot be at $p$ without both $p$ and $v$ being faulty.
Thus, in any configuration where $v \neq p$, we can always eliminate at least one of $p$ or $q$.
\textbf{Case~3:} $p \notin S$ and $q \notin S$.
By property~\toroman{4}, $\exists x,y,z \in (N(p) \cap S) \triangle (N(q) \cap S)$ with $x \neq y$, $y \neq z$, and $z \neq x$; without loss of generality let $|\{x,y,z\} \cap N(p)| > |\{x,y,z\} \cap N(q)|$.
If $v \notin N[p] \cup N[q]$, then for the intruder to be at $p$ or $q$, $v$ must be faulty, implying every detector in $N[p] \cup N[q]$ is not faulty.
Thus, if $N(p) \cap S \not\subseteq A$, then the intruder cannot be at $p$; otherwise we assume $N(p) \cap S \subseteq A$.
Similarly, if $N(q) \cap S \not\subseteq A$, then the intruder cannot be at $q$; otherwise we assume $N(q) \cap S \subseteq A$.
Therefore, $\{x,y,z\} \subseteq A$.
By hypothesis, $|\{x,y,z\} \cap N(p)| \ge 2$, which implies that the intruder cannot be at $q$.
Thus, if $v \notin N[p] \cup N[q]$, then we can always eliminate $p$ or $q$; otherwise, we assume $v \in N[p] \cup N[q]$.
Suppose $v \in N(p)$.
If $(\{x,y,z\} \cap N(p)) - (A \cup \{v\}) \neq \varnothing$, then the intruder cannot be at $p$; otherwise, we assume $(\{x,y,z\} \cap N(p)) - (A \cup \{v\}) = \varnothing$, implying $(\{x,y,z\} \cap N(p)) \subseteq (A \cup \{v\})$.
Because $\{x,y,z\} \cap N(p) \ge 2$ by hypothesis, we find that the intruder cannot be at $q$, as otherwise we would have multiple errors.
Thus, if $v \in N(p)$ then we can always eliminate at least one of $p$ or $q$; next, we consider the other possibility: when $v \in N(q) - N(p)$.
If $\{x,y,z\} \cap N(p) \cap A \neq \varnothing$ then the intruder cannot be at $q$, as otherwise there would be multiple errors; thus, we assume $\{x,y,z\} \cap N(p) \cap A = \varnothing$.
Because $|\{x,y,z\} \cap N(p)| \ge 2$ by hypothesis, there must be at least two detectors in $N(p)$ which transmit 0, so the intruder cannot be at $p$.
Therefore, we find that in any configuration we can always eliminate at least one of $p$ or $q$.
Finally, suppose two distinct detectors $u,v \in S$ transmit 2.
By hypothesis, there can be at most one false positive, so we know that an intruder is present.
We assume there can be at most one intruder; thus, the intruder is either at $v$ and $u$ incorrectly transmitted 2 or the intruder is at $u$ and $v$ incorrectly transmitted 2.
By property~\toroman{2}, $\exists w \in (N(v) \cap S) \triangle (N(u) \cap S) - \{v,u\}$; $w \neq u$, $w \neq v$, and $w$ is adjacent to $u$ or $v$, but not both.
Detector $w$ will correctly transmit either 0 or 1; if $w \in N(v) \oplus w \in A$, then the intruder is at $u$; otherwise the intruder is at $v$.
Thus, the intruder is found.

The previous cases have established that, given any arbitrary, valid detector transmission pattern, we can take any distinct pairs of vertices in the graph and eliminate one or both of them as possible intruder locations.
Via process of elimination, we will arrive at either no valid location, meaning there is no intruder, or a single vertex, the intruder location.
Additionally, this is done assuming at most one intruder and at most one false positive or false negative; thus, $S$ is an ERR:LD set, completing the proof.
\end{proof}

\FloatBarrier

\end{document}